\documentclass[mathleft,fleqn,%
]{an}
\usepackage{graphicx}
\usepackage{times}
\usepackage{bm}
\usepackage{color}
\usepackage{gensymb}
\usepackage{natbib}

\begin{document}
\sloppy

\Pagespan{1}{}
\Yearpublication{2006}%
\Yearsubmission{2005}%
\Month{11}%
\Volume{999}%
\Issue{88}%

\newcommand{\EQ}{\begin{equation}}
\newcommand{\EN}{\end{equation}}
\newcommand{\eq}[1]{(\ref{#1})}
\newcommand{\Eq}[1]{equation~(\ref{#1})}
\newcommand{\Eqs}[2]{equation~(\ref{#1}) and (\ref{#2})}
\newcommand{\Fig}[1]{Fig.~\ref{#1}}
\newcommand{\Figs}[2]{Figs.~\ref{#1} and \ref{#2}}
\newcommand{\Tab}[1]{Table~\ref{#1}}
\newcommand{\bra}[1]{\langle #1\rangle}
\newcommand{\fluc}[1]{#1^\prime}
\newcommand{\vari}[1]{#1^{\rm v}}
\newcommand{\mean}[1]{\overline#1}
\newcommand{\meanrho}{\overline{\rho}}
\newcommand{\meanPhi}{\overline{\Phi}}
\newcommand{\tildeFFFF}{\tilde{\mbox{\boldmath ${\cal F}$}}{}}{}
\newcommand{\hatFFFF}{\hat{\mbox{\boldmath ${\cal F}$}}{}}{}
\newcommand{\meanFFFF}{\overline{\mbox{\boldmath ${\cal F}$}}{}}{}
\newcommand{\meanemf}{\overline{\cal E} {}}
\newcommand{\meanAAAA}{\overline{\mbox{\boldmath ${\mathsf A}$}} {}}
\newcommand{\meanSSSS}{\overline{\mbox{\boldmath ${\mathsf S}$}} {}}
\newcommand{\meanAAA}{\overline{\mathsf{A}}}
\newcommand{\meanSSS}{\overline{\mathsf{S}}}
\newcommand{\meanuu}{\overline{\mbox{\boldmath $u$}}{}}{}
\newcommand{\meanoo}{\overline{\mbox{\boldmath $\omega$}}{}}{}
\newcommand{\meanEMF}{\overline{\mbox{\boldmath ${\cal E}$}}{}}{}
\newcommand{\meanEEEE}{\overline{\mbox{\boldmath ${\cal E}$}}{}}{}
\newcommand{\meanuxB}{\overline{\mbox{\boldmath $\delta u\times \delta B$}}{}}{}
\newcommand{\meanJB}{\overline{\mbox{\boldmath $J\cdot B$}}{}}{}
\newcommand{\meanAB}{\overline{\mbox{\boldmath $A\cdot B$}}{}}{}
\newcommand{\meanjb}{\overline{\mbox{\boldmath $j\cdot b$}}{}}{}
\newcommand{\meanAA}{\overline{\mbox{\boldmath $A$}}{}}{}
\newcommand{\meanBB}{\overline{\mbox{\boldmath $B$}}{}}{}
\newcommand{\meanEE}{\overline{\mbox{\boldmath $E$}}{}}{}
\newcommand{\meanFF}{\overline{\mbox{\boldmath $F$}}{}}{}
\newcommand{\meanFFf}{\overline{\mbox{\boldmath $F$}}_{\rm f}{}}{}
\newcommand{\meanFFm}{\overline{\mbox{\boldmath $F$}}_{\rm m}{}}{}
\newcommand{\hatFF}{\hat{\mbox{\boldmath $F$}}{}}{}
\newcommand{\meanGG}{\overline{\mbox{\boldmath $G$}}{}}{}
\newcommand{\tildeGG}{\tilde{\mbox{\boldmath $G$}}{}}{}
\newcommand{\meanJJ}{\overline{\mbox{\boldmath $J$}}{}}{}
\newcommand{\meanUU}{\overline{\bm{U}}}
\newcommand{\meanVV}{\overline{\bm{V}}}
\newcommand{\meanWW}{\overline{\mbox{\boldmath $W$}}{}}{}
\newcommand{\hatWW}{\hat{\mbox{\boldmath $W$}}{}}{}
\newcommand{\meanQQ}{\overline{\mbox{\boldmath $Q$}}{}}{}
\newcommand{\meanA}{\overline{A}}
\newcommand{\meanB}{\overline{B}}
\newcommand{\meanb}{\tilde{b}}
\newcommand{\meanj}{\tilde{j}}
\newcommand{\meanC}{\overline{C}}
\newcommand{\meanF}{\overline{F}}
\newcommand{\meanFm}{\overline{F}_{\rm m}}
\newcommand{\meanFf}{\overline{F}_{\rm f}}
\newcommand{\meanh}{\overline{h}}
\newcommand{\meanhm}{\overline{h}_{\rm m}}
\newcommand{\meanhf}{\overline{h}_{\rm f}}
\newcommand{\meanG}{\overline{G}}
\newcommand{\meanGGG}{\overline{\cal G}}
\newcommand{\meanH}{\overline{H}}
\newcommand{\meanU}{\overline{U}}
\newcommand{\meanR}{\overline{\rho}}
\newcommand{\meanJ}{\overline{J}}
\newcommand{\means}{\overline{s}}
\newcommand{\meanp}{\overline{p}}
\newcommand{\meanS}{\overline{S}}
\newcommand{\meanT}{\overline{T}}
\newcommand{\meanW}{\overline{W}}
\newcommand{\meanQ}{\overline{Q}}
\newcommand{\meanEEE}{\overline{\cal E}}
\newcommand{\meanFFF}{\overline{\cal F}}
\newcommand{\flucbb}{\mbox{\boldmath $b^\prime$}{}}{}
\newcommand{\flucuu}{\mbox{\boldmath $u^\prime$}{}}{}
\newcommand{\flucbbz}{\mbox{\boldmath $b^\prime_0$}{}}{}
\newcommand{\flucuuz}{\mbox{\boldmath $u^\prime_0$}{}}{}
\newcommand{\hatAA}{\hat{\bm{A}}}
\newcommand{\hatBB}{\hat{\bm{B}}}
\newcommand{\hatJJ}{\hat{\mbox{\boldmath $J$}}{}}{}
\newcommand{\hatOO}{\hat{\bm{\Omega}}}
\newcommand{\hatEMF}{\hat{\mbox{\boldmath ${\cal E}$}}{}}{}
\newcommand{\emf}{{\cal E}}{}
\newcommand{\hatA}{\hat{A}}
\newcommand{\hatB}{\hat{B}}
\newcommand{\hatJ}{\hat{J}}
\newcommand{\hatU}{\hat{U}}
\newcommand{\alphaK}{\alpha_{\rm K}}
\newcommand{\alphaM}{\alpha_{\rm M}}
\newcommand{\alpharr}{\alpha_{rr}}
\newcommand{\alphatt}{\alpha_{\theta\theta}}
\newcommand{\alphapp}{\alpha_{\phi\phi}}
\newcommand{\meanBBp}{\meanBB_{\rm pol}}
\newcommand{\meanUUt}{\meanUU_{\rm tor}}
\newcommand{\meanUUp}{\meanUU_{\rm pol}}
\newcommand{\meanBBt}{\meanBB_{\rm tor}}
\newcommand{\bbmBB}{\flucbb_{\scriptsize\meanBB\normalsize}}
\newcommand{\uumBB}{\flucuu_{\scriptsize\meanBB \normalsize}}
%
%
\newcommand{\teps}{\tilde{\epsilon} {}}
\newcommand{\Ot}{\tilde{\Omega}}
\newcommand{\zh}{\hat{z}}
\newcommand{\PC}{{\sc Pencil Code}~}
%
%
\newcommand{\pphi}{\hat{\bm{\phi}}}
\newcommand{\ppom}{\bm{\hat{\varpi}}}
\newcommand{\eee}{\hat{\mbox{\boldmath $e$}} {}}
\newcommand{\nnn}{\hat{\mbox{\boldmath $n$}} {}}
\newcommand{\rrr}{\hat{\mbox{\boldmath $r$}} {}}
\newcommand{\vvv}{\hat{\mbox{\boldmath $v$}} {}}
\newcommand{\xxx}{\hat{\mbox{\boldmath $x$}} {}}
\newcommand{\yyy}{\hat{\mbox{\boldmath $y$}} {}}
\newcommand{\zzz}{\hat{\mbox{\boldmath $z$}} {}}
\newcommand{\ttt}{\hat{\mbox{\boldmath $\theta$}} {}}
\newcommand{\OOO}{\hat{\mbox{\boldmath $\Omega$}} {}}
\newcommand{\ooo}{\hat{\mbox{\boldmath $\omega$}} {}}
\newcommand{\BBBB}{\hat{\mbox{\boldmath $B$}} {}}
\newcommand{\meanBBhat}{\hat{\overline{\bm{B}}}}
\newcommand{\meanJJhat}{\hat{\overline{\bm{J}}}}
\newcommand{\meanBhat}{\hat{\overline{B}}}
\newcommand{\meanJhat}{\hat{\overline{J}}}
\newcommand{\Bhat}{\hat{B}}
\newcommand{\Bh}{\hat{B}}
%
%
\newcommand{\nullvector}{{\bf0}}
\newcommand{\gggg}{\mbox{\boldmath $g$} {}}
\newcommand{\ddd}{\mbox{\boldmath $d$} {}}
\newcommand{\rr}{\mbox{\boldmath $r$} {}}
\newcommand{\yy}{\mbox{\boldmath $y$} {}}
\newcommand{\zz}{\mbox{\boldmath $z$} {}}
\newcommand{\vv}{\mbox{\boldmath $v$} {}}
\newcommand{\ww}{\mbox{\boldmath $w$} {}}
\newcommand{\mm}{\mbox{\boldmath $m$} {}}
\newcommand{\PP}{\mbox{\boldmath $P$} {}}
\newcommand{\bp}{\mbox{\boldmath $p$} {}}
\newcommand{\II}{\mbox{\boldmath $I$} {}}
\newcommand{\RR}{\mbox{\boldmath $R$} {}}
\newcommand{\kk}{\bm{k}}
\newcommand{\pp}{\bm{p}}
\newcommand{\qq}{\bm{q}}
\newcommand{\xx}{\bm{x}}
\newcommand{\XX}{\bm{X}}
\newcommand{\VV}{\bm{V}}
\newcommand{\KK}{\mbox{\boldmath $K$} {}}
\newcommand{\uu}{\mbox{\boldmath $u$} {}}
\newcommand{\UU}{\mbox{\boldmath $U$} {}}
\newcommand{\sss}{\mbox{\boldmath $s$} {}}
\newcommand{\ssss}{\mbox{\boldmath $\xi$} {}}
\newcommand{\bb}{\mbox{\boldmath $b$} {}}
\newcommand{\BB}{\mbox{\boldmath $B$} {}}
\newcommand{\EE}{\mbox{\boldmath $E$} {}}
\newcommand{\jj}{\mbox{\boldmath $j$} {}}
\newcommand{\JJ}{\mbox{\boldmath $J$} {}}
\newcommand{\SSS}{\mbox{\boldmath $S$} {}}
\newcommand{\AAA}{\mbox{\boldmath $A$} {}}
\newcommand{\aaaa}{\mbox{\boldmath $a$} {}}
\newcommand{\ee}{\mbox{\boldmath $e$} {}}
\newcommand{\nn}{\mbox{\boldmath $n$} {}}
\newcommand{\ff}{\mbox{\boldmath $f$} {}}
\newcommand{\hh}{\mbox{\boldmath $h$} {}}
\newcommand{\FF}{\mbox{\boldmath $F$} {}}
\newcommand{\EEE}{\mbox{\boldmath ${\cal E}$} {}}
\newcommand{\FFF}{\mbox{\boldmath ${\cal F}$} {}}
\newcommand{\TT}{{\bm{T}}}
\newcommand{\MM}{\mbox{\boldmath $M$} {}}
\newcommand{\GG}{\mbox{\boldmath $G$} {}}
\newcommand{\WW}{\mbox{\boldmath $W$} {}}
\newcommand{\QQ}{\mbox{\boldmath $Q$} {}}
\newcommand{\grav}{\mbox{\boldmath $g$} {}}
\newcommand{\nab}{\mbox{\boldmath $\nabla$} {}}
\newcommand{\OO}{\bm{\Omega}}
\newcommand{\oo}{\mbox{\boldmath $\omega$} {}}
\newcommand{\pom}{\mbox{\boldmath $\varpi$} {}}
\newcommand{\ttau}{\mbox{\boldmath $\tau$} {}}
\newcommand{\LL}{\mbox{\boldmath $\Lambda$} {}}
\newcommand{\mmu}{\mbox{\boldmath $\mu$} {}}
\newcommand{\ddelta}{\mbox{\boldmath $\delta$} {}}
\newcommand{\ggamma}{\mbox{\boldmath $\gamma$} {}}
\newcommand{\aalpha}{\mbox{\boldmath $\alpha$} {}}
\newcommand{\bbeta}{\mbox{\boldmath $\beta$} {}}
\newcommand{\kkappa}{\mbox{\boldmath $\kappa$} {}}
\newcommand{\llambda}{\mbox{\boldmath $\lambda$} {}}
\newcommand{\pomega}{\mbox{\boldmath $\varpi$} {}}
%
%
\newcommand{\DDDD}{\mbox{\boldmath ${\sf D}$} {}}
\newcommand{\IIII}{\mbox{\boldmath ${\sf I}$} {}}
\newcommand{\LLLL}{\mbox{\boldmath ${\sf L}$} {}}
\newcommand{\MMMM}{\mbox{\boldmath ${\sf M}$} {}}
\newcommand{\NNNN}{\mbox{\boldmath ${\sf N}$} {}}
\newcommand{\PPPP}{\mbox{\boldmath ${\sf P}$} {}}
\newcommand{\QQQQ}{\mbox{\boldmath ${\sf Q}$} {}}
\newcommand{\RRRR}{\mbox{\boldmath ${\sf R}$} {}}
\newcommand{\SSSS}{\mbox{\boldmath ${\sf S}$} {}}
\newcommand{\BBBBB}{\mbox{\boldmath ${\sf B}$} {}}
\newcommand{\tAAAA}{\tilde{\mbox{\boldmath ${\sf A}$}} {}}
\newcommand{\tDDDD}{\tilde{\mbox{\boldmath ${\sf D}$}} {}}
\newcommand{\tRRRR}{\tilde{\mbox{\boldmath ${\sf R}$}} {}}
\newcommand{\tQQQQ}{\tilde{\mbox{\boldmath ${\sf Q}$}} {}}
\newcommand{\AAAA}{\mbox{\boldmath ${\cal A}$} {}}
\newcommand{\BBB}{\mbox{\boldmath ${\cal B}$} {}}
\newcommand{\EMF}{\mbox{\boldmath ${\cal E}$} {}}
\newcommand{\GGG}{\mbox{\boldmath ${\cal G}$} {}}
\newcommand{\HHH}{\mbox{\boldmath ${\cal H}$} {}}
\newcommand{\QQQ}{\mbox{\boldmath ${\cal Q}$} {}}
\newcommand{\GGGG}{{\bf G} {}}
%
%
\newcommand{\ii}{{\rm i}}
\newcommand{\erf}{{\rm erf}}
\newcommand{\grad}{{\rm grad} \, {}}
\newcommand{\curl}{{\rm curl} \, {}}
\newcommand{\dive}{{\rm div}  \, {}}
\newcommand{\Dive}{{\rm Div}  \, {}}
\newcommand{\diag}{{\rm diag}  \, {}}
\newcommand{\sgn}{{\rm sgn}  \, {}}
\newcommand{\DD}{{\rm D} {}}
\newcommand{\DDD}{{\cal D} {}}
\newcommand{\dd}{{\rm d} {}}
\newcommand{\dV}{\,{\rm d}V {}}
\newcommand{\dS}{\,{\rm d}{{\bm{S}}} {}}
\newcommand{\const}{{\rm const}  {}}
\newcommand{\CR}{{\rm CR}}

\def\Ta{\mbox{\rm Ta}}
\def\Ra{\mbox{\rm Ra}}
\def\Ma{\mbox{\rm Ma}}
\def\Co{\mbox{\rm Co}}
\def\Sh{\mbox{\rm Sh}}
\def\St{\mbox{\rm St}}
\def\Roo{\mbox{\rm Ro}^{-1}}
\def\Rooo{\mbox{\rm Ro}^{-2}}
\def\PrSGS{\mbox{\rm Pr}_{\rm SGS}}
\def\Pr{\mbox{\rm Pr}}
\def\Sc{\mbox{\rm Sc}}
\def\Tr{\mbox{\rm Tr}}
\def\Pm{\mbox{\rm Pr}_{\rm M}}
\def\Rm{\mbox{\rm Re}_{\rm M}}
\def\Rmc{R_{\rm m,{\rm crit}}}
\def\Rey{\mbox{\rm Re}}
\newcommand{\Rc}{{{R_{C}}}}
\def\Pe{\mbox{\rm Pe}}
\def\Co{\mbox{\rm Co}}
\def\Lu{\mbox{\rm Lu}}
\def\csz{c_{\rm s0}}
\def\cs{c_{\rm s}}
\def\pt{p_{\rm t}}
\def\ptz{p_{\rm t0}}
\def\vA{v_{\rm A}}
\def\hf{h_{\rm f}}
\def\hm{h_{\rm m}}
\def\kmean{k_{\rm m}}
\def\lf{l_{\rm f}}
\def\kf{k_{\rm f}}
\def\Ff{F_{\rm f}}
\def\Fm{F_{\rm m}}
\def\Hf{H_{\rm f}}
\def\Hm{H_{\rm m}}
\def\vArms{v_{\rm A,rms}}
\def\Brms{B_{\rm rms}}
\def\Jrms{J_{\rm rms}}
\def\Urms{U_{\rm rms}}
\def\urms{u_{\rm rms}}
\def\urmsp{u^\prime_{\rm rms}}
\def\uref{u_{\rm ref}}
\def\kappaBB{\kappa_{\rm BB}}
\def\kappaB{\kappa_{\rm B}}
\def\kappaOO{\kappa_{\Omega\Omega}}
\def\kappaO{\kappa_{\Omega}}
\def\kappah{\kappa_{\rm h}}
\def\kappat{\kappa_{\rm t}}
\def\kappatz{\kappa_{\rm t0}}
\def\nut{\nu_{\rm t}}
\def\etat{\eta_{\rm t}}
\def\etatz{\eta_{\rm t0}}
\def\etaTz{\eta_{\rm T0}}
\def\nutz{\nu_{\rm t0}}
\def\etaT{\eta_{\rm T}}
\def\mut{\mu_{\rm t}}
\def\muT{\mu_{\rm T}}
\def\uT{\mu{\rm T}}
\def\BBeq{|\BB|/B_{\rm eq}}
\def\Beq{B_{\rm eq}}
\def\Btt{\overline{B}{}_\phi^{\rm rms}}
\def\Brt{\overline{B}{}_r^{\rm rms}}
\def\Bpt{\overline{B}{}_\theta^{\rm rms}}


\def\onethird{{\textstyle{1\over3}}}
\def\onehalf{{\textstyle{1\over2}}}

\newcommand{\yprl}[3]{ #1, {PRL,} {#2}, #3}

\title{Long-term variations of turbulent transport coefficients in a solar-like convective dynamo simulation}

\author{F. A. Gent\inst{1}\fnmsep\thanks{Corresponding author: frederick.gent@aalto.fi}
    \and
    Maarit J. K\"apyl\"a\inst{2,1}
    \and
    J. Warnecke\inst{2}}
\titlerunning{Long-term variation of turbulent transport coefficients}
\authorrunning{F. A. Gent}
\institute{
ReSoLVE Centre of Excellence, Department of Computer Science, PO BOX 15400, 00760 Aalto University, Helsinki, Finland
\and 
Max Planck Institute for Solar System Research, Justus-von-Liebig-Weg 3, 37707 G\"ottingen, Germany}

\received{2017 Sep 10}
\accepted{2017 Aug 22}


\abstract{%
The Sun, aside from its eleven year sunspot cycle is additionally subject to
long term variation in its activity.
In this work we analyse a solar-like convective dynamo simulation,
containing approximately 60 magnetic cycles, exhibiting equatorward propagation
of the magnetic field, multiple frequencies, and irregular variability,
including a missed cycle and complex parity transitions between dipolar and
  quadrupolar modes.
We compute the turbulent transport coefficients, describing the effects of the turbulent velocity field
  on the mean magnetic field, using the test-field method.
The test-field analysis provides a plausible explanation of the missing cycle in terms of
the reduction of $\alpha_{\phi\phi}$ \emph{in advance of} the reduced surface
activity, and enhanced downward turbulent pumping \emph{during} the event to
confine some of the magnetic field at the bottom of the convection zone, 
where local maximum of magnetic energy is observed during the event.
At the same time, however, a quenching of the turbulent magnetic
diffusivities is observed, albeit differently distributed in depth compared to the other transport
  coefficients. Therefore, dedicated mean-field modelling is required for verification.
  }

\maketitle

\section{Introduction}\label{sect:intro}

Simulations of stellar global convection with
spontaneous dynamo action have become abundant during
the recent years thanks to the ever increasing computational resources.
In recent years we have witnessed the emergence of realistic solar-like
dynamo solutions, where the large-scale magnetic field shows polarity
reversals and  equatorward propagation \citep[e.g.][]{GCS10,KMB12,ABMT15} with the important
realisation that the presence of a shear layer at the bottom of the
convection zone (CZ) is not important in obtaining such solutions.
In addition to the basic solar cycle, some models also
capture multiple dynamo modes in one and the same simulation
\citep{KKOBWK16,BSCC16}, and also exhibit irregular behaviour
resembling, at least to some extent, grand minima
\citep{ABMT15,KKOBWK16,KKOWB16}.

It turns out that the equatorward migration direction of the large-scale
magnetic field in these simulations can be well
explained with the propagation direction of an $\alpha\Omega$ dynamo
wave following the Parker-Yoshimura sign rule \citep{Pa55b,Yo75} as
shown in several studies \citep{WKKB14,WKKB16,GSGKM16,KKOBWK16, KKOWB16}
using the kinetic and magnetic helicities as proxies to compute the
$\alpha$ effect.
However, this result crucially relies on the sign and distribution of the
$\alpha$ effect and therefore a measurement of it rather than
estimations using proxies would be desirable \citep{WRKKB16}.

It must be regarded as a major breakthrough that solar-like dynamo
solutions can now be obtained from global convection simulations.
This in itself, however, does not as yet reveal the internal workings of the
solar dynamo.
The next challenge is to carefully analyze such simulation
results to pinpoint how various possible mechanisms, namely
differential rotation, meridional circulation, and turbulent effects,
affect the evolution of the mean magnetic field. This is another
non-trivial task, especially where the measurement of the turbulent
effects is concerned.

One method to determine the turbulent transport coefficients and
therefore the effect of turbulent flow on the evolution of the mean magnetic
field,
is the \emph{test-field method} \citep{SRSRC05,SRSRC07}.
This method makes use of the mean-field approximation, and
utilises linearly independent test-fields, which do not backreact
on the dynamo solution, but are merely used to measure the turbulent
transport coefficients.
We assume that the large-scale field is axisymmetric about the polar
axis and can therefore be described by a mean field averaged over
the azimuthal direction satisfying the Reynolds averaging rules,
required for the mean-field approximation to hold.

This method has succesfully been applied to
simulations of planetary interiors
\citep[e.g.][]{SRSRC05,SRSRC07,SPD12}, and various other
setups \citep[see e.g.][and references therein]{BCDHKR10}. 
Recently it was applied, for the first time, to solar-like convective
dynamo solutions by \cite{WRKKB16}.
The general conclusion is that turbulent effects play an important role in
the dynamics and evolution of the mean magnetic field in these simulations.
More specifically, the authors found that the components of
$\aalpha$ tensor, and
therefore the $\alpha$ effect, are not well described by the isotropic
SOCA expression based on kinetic helicity.
Furthermore, the turbulent pumping significantly alters the effective
meridional circulation acting on the magnetic field.
Global convection simulations challenged the flux-transport
dynamo paradigm already before, as multiple circulation cells
are more a rule than an exception \citep[see e.g.][]{KKB14},
while the one-cell circulation
important for these dynamo models seems unlikely. The new results,
in practise, imply that even though the meridional flow measurement
accuracy improved to provide a reliable determination of the pattern,
the turbulent effects would still elude observations, making the
measurement of the effective meridional circulation pattern relevant
for the mean magnetic field impossible.
Moreover, all the turbulent transport coefficients show variation along the magnetic cycle, and
additionally strong random variation, the former of which can play an
important role in the saturation process of the dynamo.
\cite{WRKKB16} additionally found that coefficients calculated with
methods based on linear regression as used in
e.g.,\,\cite{RCGBS11} and \cite{ABMT15} can be incorrect and therefore lead to
misleading conclusions for the dynamo mechanism, i.e\ the wrong
progation direction of the dynamo wave.

The solar-like dynamo solutions have been integrated over timescales of the order
of a hundred magnetic cycles, and have revealed interesting long-term
phenomena (a longer cycle, reminiscent of the Gleissberg cycle, and
irregular behaviour,
including the disappearance of surface magnetic fields), as reported in \cite{KKOBWK16}.
Simulations evolving an equivalent time span, but over fewer 
magnetic cycles, and without significant irregularity, have also been
produced by \cite{PC14} and \cite{NCP14}.
In the study of \cite{KKOBWK16}, no evident reason due to differential rotation
nor meridional circulation could be found for the irregular behaviour, while
the turbulent transport coefficients were not determined.
All the interesting behaviour was, however, in one way or another related
to the properties of and competition between the different dynamo modes present in the system.

Therefore, it is of great interest to understand why and when different
types of dynamo modes become excited in such a system.
For this purpose, the turbulent transport coefficients need to be determined, 
which is the goal of this paper.
We study a dynamo solution that was restarted from a snapshot 
of the \cite{KKOBWK16} simulation, but varying slightly the boundary
conditions and including the test field module, somewhat altering
the length of timestep. 
Consequently, we obtain a statistically equivalent dynamo solution, such that
it differs only locally in time.
We concentrate our analysis on understanding the epochs
of varying parity of the basic magnetic cycle,
and one irregular epoch when the surface magnetic field vanished from the
northern hemisphere for a time scale of the order of one cycle.

\section{Model and setup}\label{sect:mod}

The simulation setup is the same as in \cite{KKOBWK16} and a detailed
description of the general model can be found in \cite{KMCWB13} and
will not be repeated here.
We model the solar convection zone as a spherical wedge in spherical
polar
coordinates: $0.7\,R \le r\le 1\,R$ in radius, $15^\circ \le \theta \le
165^\circ$ in latitude and $0^\circ \le \phi \le 90^\circ$ in azimuth,
where $R$ is the solar radius.
We solve the compressible magnetohydrodynamic equations for the
plasma density $\rho$, velocity $\uu$, specific entropy $s$ and the
magnetic vector potential $\AAA$, with $\BB=\nab\times\AAA$, 
assuming an ideal gas for the equation of state.
We include the effects of gravity and rotation on the fluid evolution.
The details of the setup, the exact form of the equations and the
initial conditions can be found in 
\cite{KMCWB13,KKOBWK16} and \cite{WKKB14}. 

The spherical wedge is assumed to be periodic along the azimuthal direction.
The boundary conditions
at the inner radial and both latitudinal boundaries satisfy a perfect conductor, and a radial
magnetic field condition at the outer radial boundary.
The present setup includes a small correction to the radial perfect
conductor condition in comparison with \cite{KKOBWK16}.
This adds an extra term that must be included when using
spherical rather than Cartesian coordinates to fullfill vanishing
tangential currents.
However, the inclusion of this term does not affect the overall statistics of the dynamo
solution.
The flow obeys a stress-free condition at all radial and
latitudinal boundaries.
For entropy the lower radial boundary condition is described by
a constant heat flux into the wedge, and at the upper boundary the
temperature obeys a blackbody condition.
The thermodynamic quantities have zero derivatives leading to
vanishing energy fluxes at the latitudinal boundaries.

Our model can be characterized with fluid and magnetic Prandtl
numbers
\[\PrSGS=\frac{\nu}{\chi^{\rm m}_{\rm SGS}}=1,\quad
\Pm   =\frac{\nu}{\eta}=1,\]
Reynolds numbers
\[\Rey=\frac{\urms}{\nu\kf}=29,\quad  \Rm=\frac{\urms}{\eta\kf}=29 \]
and Coriolis number
\[ \Co=2\frac{\Omega_0}{\urms\kf}=9.5,\] where $\nu$
is the kinematic viscosity, $\chi^{\rm m}_{\rm SGS}$ turbulent heat
diffusion at the middle of the CZ, $\eta$ is the magnetic
diffusivity and $\Omega_0$ the rotation rate.
Here, the wavenumber of the largest
vertical scale in the CZ is $\kf=2\pi/0.3R \approx21/R$ and
\[\urms=\sqrt{\frac{3}{2}\bra{u_r^2+u_{\theta}^2}_{r\theta\phi t}}\] is 
the rms velocity without the $\phi$ component, as $u_\phi$ is dominated by
the differential rotation.
The value of the magnetic Reynolds number correspond to around
  3 times the critical value for the dynamo to be excited.
We also define the meridional distribution of the turbulent velocity
\[\urmsp(r,\theta)={\left\langle\,\overline{{\bm
        u}^{\prime\,2}}\,\right\rangle_t}{}^{\!\!1/2},\] where we
use the mean-field decomposition into mean flow $\meanUU$ and
fluctuating flow $\flucuu=\uu-\meanUU$. The mean is calculated using
the azimuthal average.
We apply a similar decomposition for the magnetic
field: $\BB=\meanBB+\flucbb$.
In this paper, we additionally use a decomposition into
time-average
$\bra{f}_t$ and variation $\vari{f}$ for such quantity $f$.
The turbulent transport coefficients presented are typically
normalized either by 
\[\alpha_0=\frac{1}{3}\urmsp \quad{\rm or} \quad 
\etatz=\frac{1}{3}\tau u^{\prime\,2}_{\rm rms}
\]
as appropriate.
An estimate of the convective turnover time is 
$\tau=H_p\alpha_{\rm MLT}/\urmsp$, where 
$H_p=-(\partial \ln\mean{p}/\partial r)^{-1}$ is the pressure scale
height and the mixing length parameter, $\alpha_{\rm MLT}$, is here set to
be $5/3$.

To present our results in physical units we invoke normalisation based
on the solar rotation rate
$\Omega_{\odot}=2.7\times10^{-6}\,$s$^{-1}$ and the density at the
bottom of convection of the Sun. For a detailed
description and discussion, relating the
simulations to real stars, we refer to \citet{KMCWB13,KKB14,KKOBWK16}
and \citet{WKKB14}.
The simulation is performed with the {\sc Pencil
  Code}\footnote{\tt{http://github.com/pencil-code/}}, which uses a
high-order finite difference method for solving the compressible
equations of MHD.

\subsection{Test-field method}

In the test-field method \citep{SRSRC05,SRSRC07,WRKKB16}, it is
assumed that the turbulent electromotive
force, $\EMF=\overline{\flucuu\times\flucbb}$,
describing the effect of turbulence on the mean magnetic field,
can be expressed as an
expansion in terms of $\meanBB$, where contributions from higher
derivatives are neglected and scale separation is assumed.
Following \cite {KR80}, this leads to 
\begin{eqnarray}
\label{eq:testpr}
\EMF&=&\aalpha\cdot\meanBB+\ggamma\times\meanBB 
-\bbeta\cdot(\nab\times\meanBB)\\\nonumber
&-&\ddelta\times(\nab\times\meanBB) 
-\kkappa \cdot(\nab\meanBB)^{(s)}.
\end{eqnarray}
Here $\aalpha$ is giving rise to the
$\alpha$ effect \citep{SKR66}.
$\ggamma$ describes changes of the mean magnetic field due to an effective
velocity \citep[or ``turbulent pumping'', e.g.,][]{OSBR02}.
$\bbeta$ characterizes the
turbulent diffusion.
$\ddelta$ enables what is known as the R\"adler effect \citep{KHR69}
and the shear-current effect.
The rank-three tensor, $\kkappa$, with the symmetric part of the derivative
tensor, $(\nab\meanBB)^{(s)}$, does not yet have an established physical
interpretation.
To calculate the coefficients of \Eq{eq:testpr}, we simultaneously solve for the
evolution of fluctuating magnetic fields using nine independent, axisymmetric
test-fields and the flow field from the global simulation.
From the nine test magnetic fields and their corresponding fluctuations,
we are then able to calculate nine electromotive forces and
invert for all coefficients in \Eq{eq:testpr}.
To avoid the eigensolution of the test-field problem becoming
unstable, we re-initialize the test fields to zero at regular time
intervals, which are typically much larger than the turnover time
and much shorter than the magnetic cycle.
For the analysis we filter out transient time moments just before
and after the resetting.
The same technique was used in \cite{WRKKB16}.
Further details and discussions about the theoretical background, the physical
interpretation and the shortcomings of the test-field method can be found in
\citet{SRSRC05,SRSRC07} and \cite{WRKKB16}.

\section{Results}\label{sect:res}

\subsection{Dynamo characteristics}\label{sect:dyn}

The simulation is continued from an original run of \cite{KKOBWK16} near 
year 25, the dynamo having reached a statistical steady state, and now with
the addition of the test-fields.
A brief interval of transitional perturbation occurs due to the above mentioned
reformulation of the lower magnetic boundary condition, and this has been
excluded from the analysis presented here.

To recap, the present simulation exhibits a solar-like differential
rotation with local minima at mid latitudes together with a
multi-cellular meridional circulation.
The main dynamo mode has a mean cycle period of around five years and shows
a clear equatorward migration.
It can be well described by a propagating dynamo wave based on the
negative shear due to the local minima at mid latitudes and a positive
$\alpha$ effect in this region \citep{KKOBWK16}.

With the inclusion of the test-fields the application of the Courant
stability condition to the additional equations requires a slight 
reduction in the typical timestep of about 20\%.
The system is highly chaotic and the difference in timestep produces a 
different realisation of the turbulence, which diverges over time from the
original simulation.
The statistical properties of the system remain consistent, but the actual
shifts in cycles, missing or weak cycles now occur at different times.
The current run, spanning over 300\,yr and amounting to roughly 60 magnetic
cycles, has yet to exhibit a grand minimum.
Such an event occurred in \cite{KKOBWK16} after about the ten first magnetic cycles, but then took more
than 80 magnetic cycles to reoccur.
The current run, however, exhibits all the co-existing three cycles, and long
term variation consistent with the original run.
During 175–-215 yr, we also observe a short epoch over time when 
both the
radial and azimuthal magnetic fields disappear from the northern hemisphere;
we will later denote this time interval as the `missing cycle'. Although
not clearly a grand minimum type event, the turbulent transport coefficients
exhibit interesting behaviour during this interval, analyzed in detail in
Section\,\ref{sect:missing}.

\begin{figure}
   \includegraphics[trim={0.0cm 0.182cm 0.05cm 0.0cm}, clip, width=\columnwidth]{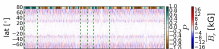}
   \includegraphics[trim={0.0cm 0.001cm 0.05cm 0.0cm}, clip, width=\columnwidth]{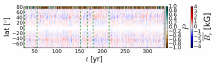}
   \includegraphics[trim={0.0cm 0.182cm 0.05cm 0.0cm}, clip, width=\columnwidth]{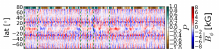}
   \includegraphics[trim={0.0cm 0.001cm 0.05cm 0.0cm}, clip, width=\columnwidth]{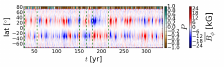}
   \caption{
   \label{fig:Braw}
   Time-latitude plot of mean magnetic field components, $\meanB_r$ 
   (upper pair) and $\meanB_\phi$ (lower pair).
   The first of each pair uses a radial slice near the surface (0.98$R$) and the second
   from near the base of the CZ (0.72$R$).
   The parity of the total magnetic field across the slice is depicted by the
   `barcode' along the top (brown-green).
   Vertical lines (green, dashed) indicate three epochs examined in 
   detail.
   }
\end{figure}

To illustrate some of these phenomena, Fig.\,\ref{fig:Braw} includes
two pairs of butterfly diagrams, i.e.\ time-latitude diagrams, for mean
magnetic field component $\meanB_r$ (upper pair) and $\meanB_\phi$ (lower) for the
surface and the bottom of the CZ.
The parity of the total field has been calculated for each slice as function of time 
and is identified by the brown-green `barcode'.
Parity is defined as 
\begin{equation}\label{eq:parity}
  P = \frac{E_{\rm even}-E_{\rm odd}}{E_{\rm even}+E_{\rm odd}},
\end{equation}
where $E_{\rm even}\,(E_{\rm odd})$ is the magnetic energy (or other 
quantity) that is equatorially symmetric (anti-symmetric), i.e.\,quadrupolar
(dipolar) with extremum 1 (-1).

\begin{figure}
   \includegraphics[trim={0.0cm 0.182cm 0.02cm 0.0cm}, clip, width=\columnwidth]{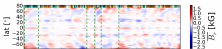}
   \includegraphics[trim={0.0cm 0.002cm 0.05cm 0.0cm}, clip, width=\columnwidth]{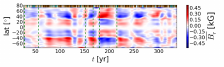}
   \includegraphics[trim={0.0cm 0.182cm 0.02cm 0.0cm}, clip, width=\columnwidth]{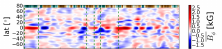}
   \includegraphics[trim={0.0cm 0.002cm 0.05cm 0.0cm}, clip, width=\columnwidth]{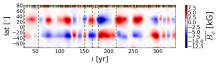}
   \caption{
     \label{fig:smBs}
   Time-latitude plot of mean magnetic field components, $\meanB_r$ 
   (upper pair) and $\meanB_\phi$ (lower pair) smoothed over
   approximately 50\,yr.
   Otherwise as Fig.\,\ref{fig:Braw}.
   }
\end{figure}

The characteristics of the field evolution are in excellent agreement with 
\citet[][see their Figs.\,1 and 2]{KKOBWK16}. 
The field near the base is consistent between this model and theirs,
indicating that 
differences in the implementation of the perfect conductor boundary condition
have not qualitatively affected the solution.
The length of the basic cycle is approximately five years, and varies somewhat
between the hemispheres \citep[for detailed analysis, see][]{OlspertIEEE}.
This is clearly evident in the surface magnetic field.
In the base of the CZ, the basic cycle is still visible, but
is clearly modulated by longer, rather irregular cycles. 
In the simulation run, with double the duration, analysed in
\cite{OlspertIEEE}, two dominant cycles
of roughly 50 and 100 years were found, and such cycles can also explain
the variations seen in the current results.
The basic cycle length fluctuates with distinct patterns in north and
south, to the
extent that over a century each hemisphere can produce a different number 
of cycles.
In addition, both magnetic field components exhibit fluctuations in strength, most
pronounced near the surface when the cycle in the northern hemisphere is
very weak around 200\,yr, an epoch to which we refer to as the
`missing cycle'.
This is coincident with an interval of strong azimuthal field at the
base, see \Fig{fig:smBs}.

The parity, plotted as barcodes over the data in Figs.\,\ref{fig:Braw} and \ref{fig:smBs},
switches sign irregulary over time, and the parity
profiles are quite distinct, perhaps independent, between the surface and the
base. 
The base field parity is consistently more negative than at the surface.
The parity switches are not obviously associated with the basic cycle, and the
long-term trends remain hidden in the raw data of
Fig.\,\ref{fig:Braw}.
To better identify them, we apply temporal smoothing using a
Gaussian kernel with a width of approximately 50\,yr.  

The smoothed magnetic field components are displayed in Fig.\,\ref{fig:smBs}. 
The long term cycle trends are now clearly visible, particulary for the 
magnetic field at the base. 
The smoothing helps to reveal the signatures of the long cycles also in
the radial field near the surface, while the surface azimuthal field
shows no evident modulation by the long cycles.
The parity of the surface field shows some correlation with the long term
cycle seen in the bottom azimuthal field: the parity switches appear
to occur often simultaneously with the changes of polarity of the base
mode.

In Figs.\,\ref{fig:Braw} and \,\ref{fig:smBs}, we also identify three epochs
bounded by dashed green vertical lines, for further analysis.
The first interval, 27--54\,yr, is identified with even parity in the surface 
magnetic field, and we shall refer to this as the `quadrupole epoch'.
The second interval, 151--166\,yr, is identified with odd parity, which we 
shall refer to as the `dipole epoch'.
These will be further examined in Section\,\ref{sect:parity}.
The third interval, 175--215\, yr, is motivated by the decreased surface magnetic field
strength seen and will be examined in Section\,\ref{sect:missing}.

\begin{figure}
   \includegraphics[trim={0.0cm 0.182cm 0.02cm 0.0cm}, clip=true, width=\columnwidth]{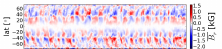}
   \includegraphics[trim={0.0cm 0.182cm 0.05cm 0.0cm}, clip=true, width=\columnwidth]{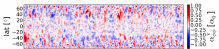}
   \includegraphics[trim={0.0cm 0.002cm 0.05cm 0.0cm}, clip=true, width=\columnwidth]{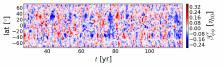}
   \caption{
   \label{fig:zoom}
   Time-latitude plot near the base of the CZ,
   depicting zoom-in of the early cycles in $\meanB_r$, $\vari{\alpha_{\phi\phi}}$ 
   and $\vari{\beta_{\phi\phi}}$. 
  Temporal Gaussian smoothing has been applied with a width of
  approximately 2.5\,yr.
  We compute time averages of the transport coefficients over the whole time
  span of the simulation, subtract this profile to obtain the residuals, and use
  the sign of the average quantity to multiply the residual to indicate
  enhancement (diminution) with respect to the average profile.
   }
\end{figure}

\subsection{Turbulent transport coefficients}\label{sect:tensors}

The derived turbulent transport coefficients are in excellent agreement with
the results of \cite{WRKKB16}: near identical to their meridional profiles, 
averaged over the
whole simulation time, and to their cyclic time variation over the basic cycle.
In the current study, we concentrate on the long term
time evolution, and zoom into epochs of interesting behaviour.

\subsubsection{Long term variability}\label{sect:vary}

In order to see the five year cycle more clearly we zoom in at the first
century of the simulation.
In Fig.\,\ref{fig:zoom} we display the butterfly diagrams for $\meanB_r$ at the 
base of the CZ and the temporal variation in the turbulent tensor coefficients
$\vari{\alpha_{\phi\phi}}$ and $\vari{\beta_{\phi\phi}}$. The 
superscript $^{\rm v}$ follows the notation convention of \citet{WRKKB16}, in
which the temporal averages are subtracted, and then the residual profile is
suitably smoothed in time, in this case
a temporal smoothing of 2.5\,yr, to retain the basic cycle.
From these diagrams it can be discerned that the cycle length of the
transport 
coefficients is about half that of the magnetic field,
due to the quadratic form of the Lorentz force
mediating the effect of the mean field on the
turbulent flow, agreeing with the results of \cite{WRKKB16}.
More significantly, both the magnetic field and the turbulent coefficients exhibit strong
fluctuations in the length and strength of their basic cycle, but also 
lower frequency modes are visible.
For example, a repeated long cycle in $\meanB_r$ is visible at 75\,yr
and 115\,yr,
with a matching equatorial quenching in $\vari{\beta_{\phi\phi}}$ at the same time,
and to a lesser extent in $\vari{\alpha_{\phi\phi}}$.

\begin{figure}
   \includegraphics[trim={0.0cm 0.182cm 0.05cm 0.0cm}, clip=true, width=\columnwidth]{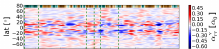}
   \includegraphics[trim={0.0cm 0.002cm 0.02cm 0.0cm}, clip=true, width=\columnwidth]{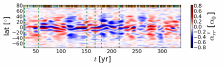}
   \includegraphics[trim={0.0cm 0.182cm 0.05cm 0.0cm}, clip=true, width=\columnwidth]{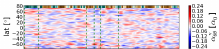}
   \includegraphics[trim={0.0cm 0.002cm 0.02cm 0.0cm}, clip=true, width=\columnwidth]{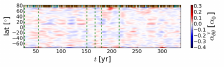}
   \includegraphics[trim={0.0cm 0.182cm 0.05cm 0.0cm}, clip=true, width=\columnwidth]{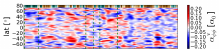}
   \includegraphics[trim={0.0cm 0.002cm 0.05cm 0.0cm}, clip=true, width=\columnwidth]{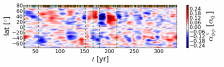}
   \caption{
     \label{fig:smas}
   Time-latitude plot of smoothed
   variation of the diagonal components of $\aalpha$.
   Pairs of slices from the surface (upper) and base (lower) of the 
   CZ. 
   The variations have been multiplied by the sign of the time average to
   identify enhancement (diminution) of the time-averaged tensor profile.
   The smoothing window is approximately 50\,yr.
   The parity of the total magnetic field across the slice is depicted by the
   `barcode' along the top (brown-green, $-1$ to $1$).
    Vertical lines (green, dashed) indicate three epochs examined in 
   detail.
   }
\end{figure}

To better expose these long term trends, we apply a 50\,yr smoothing window
to the full simulation and display the
diagonal components, $\vari{\alpha_{ii}}$,
paired from the surface and base of the CZ,
in Fig.\,\ref{fig:smas}.
Following \cite{WRKKB16}, we expect these components of $\aalpha$
to be the strongest and, therefore, most likely to influence the dynamo.
At the surface, the cycles of the long term variation in each
$\aalpha$
component have incoherent evolution across different latitudes, but the
time scale of the variations appears quite similar for each tensor coefficient.
The time scale also corresponds to the long term variation typical of
the surface magnetic field, 
 as displayed in
Fig.\,\ref{fig:smBs}, top and bottom panels.

The long term cyclic variations at the base of the CZ are
pronounced for $\alpha_{rr}$ and $\alpha_{\phi\phi}$, being
especially strong for the latter.
These variations are not, however, coherent between each component of
$\aalpha$, although within each component the cycles are
typically more
common across all latitudes than they are at the surface.
None of the $\aalpha$ components consistently match the
long term variation common to all components of the magnetic field.
For $\vari{\alpha_{\phi\phi}}$, the northern and southern hemispheres exhibit
quite distinct long term trends, with reversals occurring out of phase and 
persisting for variable periods and intensity.
Relating these to the variation in the magnetic cycles is not self-evident.

The long term trends visible in $\vari{\alpha_{ii}}$ at both the base and
the surface
are not obviously correlated with the parity trends for the total field,
depicted in the `barcode' on top of the data in the Fig.\,\ref{fig:smas}.
The parities of all the diagonal components of the $\aalpha$ tensor
are very close to being purely antisymmetric
at the surface at all times of the simulation, while much more
significant deviations from pure antisymmetry occur at the base of CZ.

An interval of interest is the untypically strong cycle in
$\vari{\alpha_{\phi\phi}}$ around 200\,yr in the northern hemisphere, which 
also coincides with the
'missing cycle' in the surface magnetic field
visible in Fig.\,\ref{fig:Braw}, panels 1 and 2.
This epoch is identified by the last pair of vertical lines, and is
investigated in Section\,\ref{sect:missing}.

\subsubsection{Parity switching}\label{sect:parity}

We now consider two intervals during the evolution of the surface magnetic field
when the parity of the total field remains either quadrupolar or dipolar for a longer period of time.
The `quadrupole epoch' is 27--54\,yr, which has even parity at the surface,
and the `dipole epoch' is
151--166\,yr. 
These are identified by the first two pairs of vertical green lines marked on
Figs.\,\ref{fig:Braw}, \ref{fig:smBs} and \ref{fig:smas}.
By contrasting the behaviour of the transport coefficients during these epochs
we seek to identify how they act on the quadrupolar and dipolar modes of the
dynamo.

\begin{figure}
   \includegraphics[trim={0.0cm 0.175cm 0.0cm 0.0cm}, clip, width=\columnwidth]{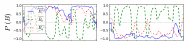}
   \includegraphics[trim={0.0cm 0.00cm 0.0cm 0.0cm}, clip, width=\columnwidth]{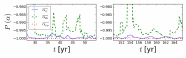}
   \includegraphics[trim={0.0cm 0.175cm 0.0cm 0.0cm}, clip, width=\columnwidth]{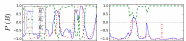}
   \includegraphics[trim={0.0cm 0.00cm 0.0cm 0.0cm}, clip, width=\columnwidth]{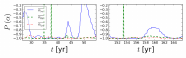}
   \caption{
   \label{fig:1Dpar}
   Parity time-series of magnetic field components and
   the diagonal components of $\aalpha$
   with 12.5\,yr smoothing window for surface (upper pair)
   and base (lower pair) of the CZ.
   The left (right) column depicts the `quadrupole (dipole) epoch' defined by
   the total magnetic field parity near the surface. 
   Note for total field parity of $\pm1$, the parity of $B_\theta$ is 
   $\mp1$.  
The spikes in particular in the lowest panels are caused by values of
alpha components being close to zero. 
   }
\end{figure}

In Fig.\,\ref{fig:1Dpar} we show the time evolution of the parity during these
epochs of each magnetic field component and
the diagonal components of $\aalpha$.
As may reasonably be anticipated the magnetic field componenents near the
surface typically match parity with the total field (first row),
although $\meanB_\theta$
most commonly least matches the total field parity.
During these intervals, the parity of the total field at the base does not
necessarily align with that at the surface.
Typically the field parity at the base throughout the whole simulation is more
consistently negative (dipolar) than at the surface.
During the `quadrupole epoch', there are significantly more
deviations from dipole at the base, than during the `dipole epoch', although
these need not be related.
At the surface the $\aalpha$ components have strong negative parity
during both dipole and
quadrupole epoch, with a modest deviation most evident in
$\alpha_{\theta\theta}$. 
Although still strongly negative, there are stronger deviations in the
$\aalpha$ components at
the base, particularly in $\alpha_{rr}$, and this is more pronounced when
the magnetic field has more quadrupolar composition.
The $\alpha_{rr}$ variations are clearly in antiphase with those of the
magnetic field, but it is difficult to establish whether the variations
in the $\aalpha$ components cause the changes in the field parity, or
vice versa.
The variations of $\alpha_{rr}$ at the base seem to be rather
tightly anti-correlated with the surface $\meanB_\theta$.
So $\alpha_{rr}$ may affect parity at the base, and seems to be
connected with the $\meanB_\theta$ parity evolution, but
the parity variations of other $\aalpha$ components seem not to be
significant.

\begin{figure}
   \includegraphics[trim={0.13cm 0.0cm 0.03cm 0.0cm}, clip=true, width=0.32\columnwidth]{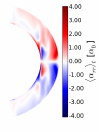}
   \includegraphics[trim={0.13cm 0.0cm 0.03cm 0.0cm}, clip=true, width=0.32\columnwidth]{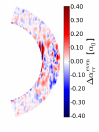}
   \includegraphics[trim={0.13cm 0.0cm 0.03cm 0.0cm}, clip=true, width=0.32\columnwidth]{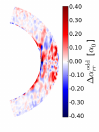}\\
   \includegraphics[trim={0.13cm 0.0cm 0.03cm 0.0cm}, clip=true, width=0.32\columnwidth]{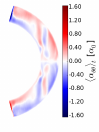}
   \includegraphics[trim={0.13cm 0.0cm 0.03cm 0.0cm}, clip=true, width=0.32\columnwidth]{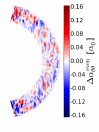}
   \includegraphics[trim={0.13cm 0.0cm 0.03cm 0.0cm}, clip=true, width=0.32\columnwidth]{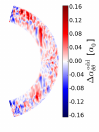}\\
   \includegraphics[trim={0.13cm 0.0cm 0.03cm 0.0cm}, clip=true, width=0.32\columnwidth]{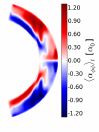}
   \includegraphics[trim={0.13cm 0.0cm 0.03cm 0.0cm}, clip=true, width=0.32\columnwidth]{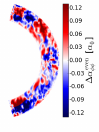}
   \includegraphics[trim={0.13cm 0.0cm 0.03cm 0.0cm}, clip=true, width=0.32\columnwidth]{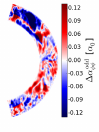}
   \caption{
   \label{fig:alp_merp}
   Meridional plot left to right of $\bra{\alpha_{ii}}_t$, averaged over all
   time, $\Delta\alpha_{ii}^{\rm even}$ averaged over the `quadrupole
   epoch' (27--54\,yr), and $\Delta\alpha_{ii}^{\rm odd}$ averaged over
   the `dipole epoch' (151--166\,yr). $\bra{\alpha_{ii}}_t$ is subtracted
   in the latter cases.
   The differences are multiplied by the sign of $\bra{\alpha_{ii}}_t$ to
   identify enhancement (diminution) relative to the total time average.
   }
\end{figure}

To examine how the transport coefficients may differ by location between
the `quadrupole' and `dipole epochs', we calculate the time averaged meridional 
profiles for each epoch and compare these to the time average for the whole
simulation.
In Fig.\,\ref{fig:alp_merp} the diagonal components of $\aalpha$ are
displayed as a time average across the total duration of
the simulation $\bra{\alpha_{ii}}_t$.
To visualize the difference between the two epochs and the time
averaged coefficients, we define
\[\Delta\alpha_{ii}^{\rm odd}=\bra{\alpha_{ii}}_{\rm
    odd}-\bra{\alpha_{ii}}_t ,
\]
\[
\Delta\alpha_{ii}^{\rm even}=\bra{\alpha_{ii}}_{\rm
    even}-\bra{\alpha_{ii}}_t,
\]
where ``even'' references to the `quadrupole epoch' and ``odd''
to the `dipole epoch'.
These are displayed in the latter two columns, multiplied by the sign of the
simulation average, so that positive (negative) $\Delta\alpha_{ii}$ indicates
enhancement (diminution) of the tensor relative to its average.

$\alpha_{rr}$ is symmetrically enhanced near the equator during the `dipole
epoch', meaning the $\alpha$ effect is amplified evenly in both 
hemispheres during this phase.
The deviations from odd parity seen during this epoch, in the lowest right
panel of Fig.\,\ref{fig:1Dpar}, are likely to be related to the narrow
low-latitude region at the base of the CZ, where $\alpha_{rr}$ is less
symmetrically enhanced or quenched.
It is less anti-symmetric during the `quadrupole epoch',
caused by the less symmetric modification profile,
and as a total, the coefficient is weakly reduced.
While $\alpha_{\theta\theta}$ does not seem to be significantly modified
comparing these two states, $\alpha_{\phi\phi}$ shows a similar pattern
as $\alpha_{rr}$: during the `dipole epoch' the quantity is symmetrically modified,
while during the `quadrupole epoch' the modification is more anti-symmetric. This results in
a tendency of the coefficient to remain close to anti-symmetric during
a dipole phase, while some deviations from pure anti-symmetry are seen
during the quadrupole phase.
These deviations, however, are smaller for $\alpha_{\phi\phi}$ than $\alpha_{rr}$,
which can also be seen from Fig.\,\ref{fig:1Dpar}.
Although there are very strong local regions of enhancement and reduction
of $\alpha_{\phi\phi}$, the net effect remains small.

\begin{figure}
   \includegraphics[trim={0.13cm 0.0cm 0.03cm 0.0cm}, clip=true, width=0.32\columnwidth]{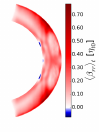}
   \includegraphics[trim={0.13cm 0.0cm 0.03cm 0.0cm}, clip=true, width=0.32\columnwidth]{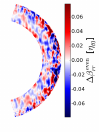}
   \includegraphics[trim={0.13cm 0.0cm 0.03cm 0.0cm}, clip=true, width=0.32\columnwidth]{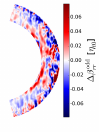}\\
   \includegraphics[trim={0.13cm 0.0cm 0.03cm 0.0cm}, clip=true, width=0.32\columnwidth]{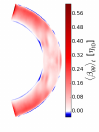}
   \includegraphics[trim={0.13cm 0.0cm 0.03cm 0.0cm}, clip=true, width=0.32\columnwidth]{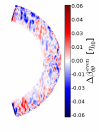}
   \includegraphics[trim={0.13cm 0.0cm 0.03cm 0.0cm}, clip=true, width=0.32\columnwidth]{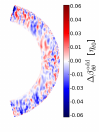}\\
   \includegraphics[trim={0.13cm 0.0cm 0.03cm 0.0cm}, clip=true, width=0.32\columnwidth]{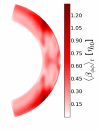}
   \includegraphics[trim={0.13cm 0.0cm 0.03cm 0.0cm}, clip=true, width=0.32\columnwidth]{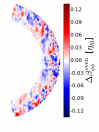}
   \includegraphics[trim={0.13cm 0.0cm 0.03cm 0.0cm}, clip=true, width=0.32\columnwidth]{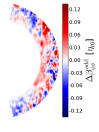}
   \caption{
   \label{fig:bet_merp}
The same as \Fig{fig:alp_merp}, but for the diagonal $\bbeta$
components.
   }
\end{figure}

The same calculations are illustrated in Fig.\,\ref{fig:bet_merp} for the
turbulent diffusion tensor $\bbeta$, which, when averaged over whole
time, are typically 
positive and symmetric about the equator. Very small negative regions appear
close to the equator at the base for $\beta_{rr}$ and
$\beta_{\theta\theta}$, and for the latter also in a narrow layer near
the surface. 
In \cite{WRKKB16}, the authors also performed a detailed study, using all the $\bbeta$
tensor coefficients,
on the possibility for negative magnetic diffusion, leading to dynamo action
instead of diffusive effects, but concluded that such effects are
probably negligibly
small in a similar system as studied here.
Similarly to the $\aalpha$ components, the $\bbeta$ components
are more symmetrically enhanced/reduced during the odd phase, while the modification
is more antisymmetric during the even phase. The overall reduction/enhancement,
however, is rather small.
Although referring to much simpler systems than considered here 
\citet{Ra86} identified mean field solutions, in which lower dynamo numbers
(high $\beta$) tend to excite dipolar fields, while higher dynamo
numbers (low $\beta$) quadrupolar.
\begin{figure}
   \includegraphics[trim={0.13cm 0.0cm 0.03cm 0.0cm}, clip=true, width=0.32\columnwidth]{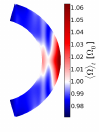}
   \includegraphics[trim={0.13cm 0.0cm 0.03cm 0.0cm}, clip=true, width=0.32\columnwidth]{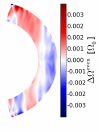}
   \includegraphics[trim={0.13cm 0.0cm 0.03cm 0.0cm}, clip=true, width=0.32\columnwidth]{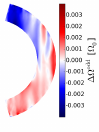}
   \caption{
   \label{fig:Omg_merp}
The same as \Fig{fig:alp_merp}, but for $\Omega$.
   }
\end{figure}

It is also informative to compare how the differential rotation profile
$\Omega=\meanU_\phi[r\sin{\theta}]^{-1}+\Omega_0$ adjusts during the
two different epochs and we define $\Delta\Omega^{\rm odd/eve}=\bra{\Omega}_{\rm odd/eve}-\bra{\Omega}_t$.
In Fig.\,\ref{fig:Omg_merp} the change $\Omega$ during each epoch indicates that during the 
`quadrupole epoch' the surface near the equator slows down and the tangent
cylinder speeds up. 
During the `dipole epoch', the equatorial surface speeds up, and the 
north (south) pole speeds up (slows down). The tangent cylinder also slows. 
Note the asymmetry evident in $\Delta\Omega$ during the `dipole epoch' 
would not arise due to the antisymmetric magnetic field, as the Lorentz force is
quadratic in $\BB$.
There must be a transient N-S gradient in magnetic energy associated with this 
epoch.
From Fig.\,\ref{fig:1Dpar} we observe this is not a pure dipole, so such 
gradients can be present.
Most importantly, the region of negative shear in the mid-latitudes, that has been found
to be instrumental in producing the equatorward propagation of the dynamo wave,
is reduced (enhanced) during the `quadrupole (dipole) epochs'. This may be the cause
of the tendency to observe less regular migration pattern and disrupted
cycles during the quadrupolar states.
The differences, however, are less than 1\%, compared to an overall spread in the 
time-averaged differential rotation rate of 10\%,
and the verification of this hypothesis will have to wait for a dedicated mean-field
investigation of the system.

\subsubsection{Missing cycle}\label{sect:missing}

\begin{figure}
   \includegraphics[trim={0.0cm 0.175cm 0.0cm 0.0cm}, clip, width=\columnwidth]{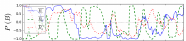}
   \includegraphics[trim={0.0cm 0.00cm 0.0cm 0.0cm}, clip, width=\columnwidth]{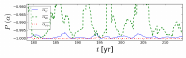}
   \includegraphics[trim={0.0cm 0.175cm 0.0cm 0.0cm}, clip, width=\columnwidth]{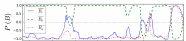}
   \includegraphics[trim={0.0cm 0.00cm 0.0cm 0.0cm}, clip, width=\columnwidth]{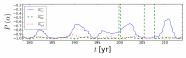}
   \caption{
   \label{fig:1Dmix}
   Parity time-series of the magnetic field components and
  diagonal components of $\aalpha$
   with 12.5\,yr smoothing window for surface (upper pair)
   and base (lower pair) of the CZ.
   Note for total field parity of $\pm1$, the parity of $B_\theta$ is 
   $\mp1$.  
   The time segment corresponds to an epoch of disruption in the surface 
   magnetic field, including a `missing cycle' in the northern hemisphere, and
   strong $\alpha_{\phi\phi}$ component near the base. 
   }
\end{figure}

In Section\,\ref{sect:vary} we identify an interval, during which a strong
variation in $\alpha_{\phi\phi}$ at the base of the convection
zone coincides with the missing cycle at the surface.
In Fig.\,\ref{fig:1Dmix} we plot the parity of the magnetic field and
diagonal
components of $\aalpha$ during this epoch 175--215\,yr near the surface and the
base.
During the `missing cycle' epoch, the surface field moves from the pre-dominantly symmetric state
into an anti-symmetric one. This means that the parity of the surface field changes from
$1$ to $-1$ during a roughly ten year time interval.
During the 'missing cycle', the base field
is highly anti-symmetric, and after surface activity is re-established, the surface field
parity starts again rapidly migrating towards 1.
Such rapid changes in the magnetic field parity are very common
throughout the simulation, but
no other such epoch is observed to produce significantly reduced
cycles.
In addition, the 'missing cycle' epoch is not preceded or
followed by any peculiar behaviour in the parities of the diagonal $\aalpha$ coefficient.
This hints towards the parity evolution being unimportant as a cause of the missed cycle.

\begin{figure}
   \centering
   \includegraphics[trim={0.0cm 0.181cm 0.05cm 0.0cm}, clip, width=\columnwidth]{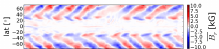}
   \includegraphics[trim={0.0cm 0.181cm 0.02cm 0.0cm}, clip, width=\columnwidth]{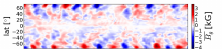}
   \includegraphics[trim={0.0cm 0.001cm 0.05cm 0.0cm}, clip, width=\columnwidth]{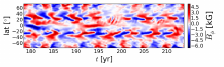}
   \includegraphics[trim={0.0cm 0.181cm 0.05cm 0.0cm}, clip, width=\columnwidth]{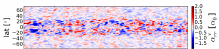}
   \includegraphics[trim={0.0cm 0.001cm 0.05cm 0.0cm}, clip, width=\columnwidth]{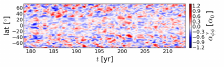}
   \includegraphics[trim={0.0cm 0.181cm 0.02cm 0.0cm}, clip, width=\columnwidth]{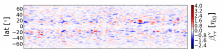}
   \includegraphics[trim={0.0cm 0.001cm 0.05cm 0.0cm}, clip, width=\columnwidth]{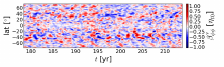}
   \includegraphics[trim={0.0cm 0.181cm 0.02cm 0.0cm}, clip, width=\columnwidth]{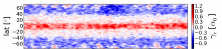}
   \includegraphics[trim={0.0cm 0.001cm 0.05cm 0.0cm}, clip, width=\columnwidth]{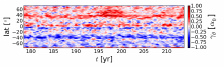}
   \caption{
   \label{fig:s_zoom}
   Zoom-in of `missing cycle' (175--215\,yr), time-latitude plots of the   
   magnetic field components and selected turbulent
  transport coefficients, with 2.5\,yr smoothing window \emph{near the
   surface}.
   For $\vari{\alpha_{ii}}$ and $\vari{\beta}_{ii}$ time averages are
   subtracted and the
   result is multiplied by the sign of the time averages.
For $\gamma_i$ no time-average is subtracted.
   }
\end{figure}

\begin{figure}
   \centering
   \includegraphics[trim={0.0cm 0.181cm 0.05cm 0.0cm}, clip, width=\columnwidth]{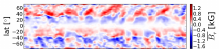}
   \includegraphics[trim={0.0cm 0.181cm 0.05cm 0.0cm}, clip, width=\columnwidth]{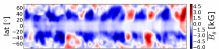}
   \includegraphics[trim={0.0cm 0.001cm 0.05cm 0.0cm}, clip, width=\columnwidth]{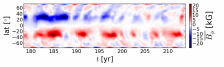}
   \includegraphics[trim={0.0cm 0.181cm 0.05cm 0.0cm}, clip, width=\columnwidth]{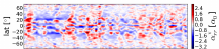}
   \includegraphics[trim={0.0cm 0.001cm 0.05cm 0.0cm}, clip, width=\columnwidth]{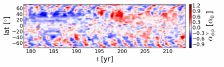}
   \includegraphics[trim={0.0cm 0.181cm 0.05cm 0.0cm}, clip, width=\columnwidth]{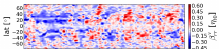}
   \includegraphics[trim={0.0cm 0.001cm 0.05cm 0.0cm}, clip, width=\columnwidth]{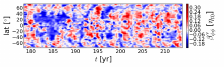}
   \includegraphics[trim={0.0cm 0.181cm 0.05cm 0.0cm}, clip, width=\columnwidth]{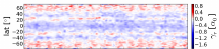}
   \includegraphics[trim={0.0cm 0.001cm 0.05cm 0.0cm}, clip, width=\columnwidth]{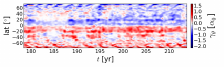}
   \caption{
     \label{fig:b_zoom}
   Zoom-in of `missing cycle' (175--215\,yr), time-latitude plot of the  
   magnetic field components and selected turbulent
   transport coefficients, with 2.5\,yr smoothing window \emph{near the base
   of the convection zone}.
   For $\vari{\alpha_{ii}}$ and $\vari{\beta_{ii}}$ time averages are subtracted
   and the
   result is multiplied by the sign of the time averages. 
For $\gamma_i$ no time-average is subtracted.
   }
\end{figure}

\begin{figure}
   \centering
   \includegraphics[trim={0.0cm 0.181cm 0.05cm 0.0cm}, clip, width=\columnwidth]{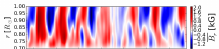}
   \includegraphics[trim={0.0cm 0.181cm 0.05cm 0.0cm}, clip, width=\columnwidth]{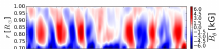}
   \includegraphics[trim={0.0cm 0.001cm 0.05cm 0.0cm}, clip, width=\columnwidth]{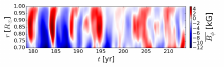}
   \includegraphics[trim={0.0cm 0.181cm 0.05cm 0.0cm}, clip, width=\columnwidth]{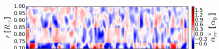}
   \includegraphics[trim={0.0cm 0.001cm 0.05cm 0.0cm}, clip, width=\columnwidth]{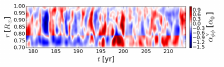}
   \includegraphics[trim={0.0cm 0.181cm 0.05cm 0.0cm}, clip, width=\columnwidth]{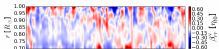}
   \includegraphics[trim={0.0cm 0.001cm 0.05cm 0.0cm}, clip, width=\columnwidth]{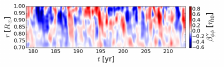}
   \includegraphics[trim={0.0cm 0.181cm 0.05cm 0.0cm}, clip, width=\columnwidth]{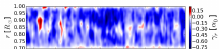}
   \includegraphics[trim={0.0cm 0.001cm 0.02cm 0.0cm}, clip, width=\columnwidth]{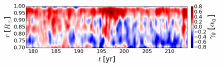}
   \caption{
   \label{fig:r_zoom}
   Zoom-in of `missing cycle' (175--215\,yr), time-radius plot of the  
   magnetic field components and selected turbulent
   transport coefficients, with 2.5\,yr smoothing window at a fixed latitude.
   For $\vari{\alpha_{ii}}$ and $\vari{\beta_{ii}}$ time averages are subtracted
   and the
   result is multiplied by the sign of the time averages.
For $\gamma_i$ no time-average is subtracted.
  The mean of ten latitudes straddling 40.6$\degree$\,N are used for the
   radial profile.
   }
\end{figure}

Explanation for this behaviour must, therefore, be sought elsewhere.
For a closer inspection of the dynamics around the 'missing cycle' in
Figs.\,\ref{fig:s_zoom}, \ref{fig:b_zoom} and \ref{fig:r_zoom}, we zoom into 
the epoch 175--215\,yr.
In Fig.\,\ref{fig:s_zoom} we display the surface magnetic field components,
and some of the more interesting
transport coefficients at the surface.
From the first three panels we can see that during the 'missing cycle', between
197 and 202\,yr,
the strength of all $\meanB$ components are reduced, particularly
those of $\meanB_r$ and
$\meanB_\phi$.
In addition, there is an enhanced high frequency variability around
20$\degree$\,N with a cycle length less than 1\,yr, most pronounced in
$\meanB_\phi$.
Such a short cycle is normally always present as the third, rather
incoherent and
poleward dynamo cycle \citep[see the analysis in][]{KKOBWK16,OlspertIEEE},
and it was observed to persist even during the grand minimum type event
seen in \cite{KKOBWK16}, consistent with the findings here.
\cite{WRKKB16} found indication that this short cycle can be
caused by a locally operating $\alpha^2$ dynamo near the surface.
Coincident with the surface field suppression, the field at the base of the CZ
obtains a local maximum in particular in $\meanB_r$ and $\meanB_\theta$, as seen
from the two top panels of Fig.\,\ref{fig:b_zoom}.
Very similar behaviour was observed by \cite{KKOBWK16} during a more
extended suppression of the surface activity, when also the enhancement of the
base magnetic field was stronger than in the current simulation.

The surface $\aalpha$ tensor components show no particularly
interesting
behavior before the 'missing cycle', but are clearly reduced during it.
$\beta_{rr}$ at the surface is somewhat enhanced before the event, and reduced after it,
but this variation is rather small.
The corresponding coefficients at the bottom, however, show more interesting
trends, that become better visible when smoothing is applied, see Fig.\ref{fig:smas}.
Especially $\alpha_{\phi\phi}$ becomes strongly reduced in the north
for several years before the event, and is enhanced after it.
A similar tendency, however, is seen also in the $\bbeta$ tensor, and
therefore
the strong reduction of $\alpha_{\phi\phi}$ appears to be compensated by
the simultaneous reduction of the diffusivities. Therefore, no great impact on
the dynamo efficiency can straightforwardly be expected, but this should be
confirmed from a mean-field analysis.

Perhaps the most interesting behavior is seen in the turbulent pumping
$\ggamma$:
during the 'missing cycle', the radial component $\gamma_r$,
directed mostly
downwards at high latitudes, is strongly enhanced there. This means that the
magnetic field is more efficiently pumped downwards during this epoch. At the same
time, similar enhancement in the latitudinal turbulent pumping $\gamma_\theta$
is seen
at a corresponding location in latitude. As the latitudinal pumping is directed
towards the equator there, this means that during the 'missing cycle' the
magnetic field
was transported more efficiently towards low latitudes. The enhancement of the
magnetic field in the bottom of the CZ is likely to be due to
the enhancements seen in the turbulent pumping.

In order to verify our conclusions based on the inspection of
time-latitude
plots, we finally inspect some time-radius plots.
We take a latitudinal mean over ten grid points straddling the approximate 
locus of the 'missing cycle' event at $40.6\degree$\,N.
Fig.\,\ref{fig:r_zoom} contains the time-radius profiles of this slice
for the magnetic field components and selected turbulent transport
coefficients.
This reveals that the region of reduced magnetic field strength permeates deep within the CZ, 
evolving earliest at the surface in $\meanB_\theta$ before 195\,yr,
a clear reduction in the magnetic field strength occurring even earlier than
that in the mid-depths for $\meanB_\theta$.

With respect to the turbulent transport, 
the reduction of $\alpha_{\phi\phi}$, $\beta_{rr}$ and $\beta_{\phi\phi}$
prior to the event itself, also
detectable in the time-latitude slices, is more clearly seen in the time-radius plots.
The latter, however, better illustrate how these effects are differently
distributed in depth.
The $\alpha_{\phi\phi}$ reduction occurs only in a narrow layer at the base of CZ,
while $\beta_{rr}$ is reduced almost at every depth except at the surface, and
$\beta_{\phi\phi}$ is reduced strongest at the topmost half of the CZ. Therefore
the implications for the dynamo efficiency are not so straightforward as
implied by the time-latitude slices.

\section{Conclusions}

In this paper we have presented the first results of an analysis
that measures
the turbulent transport coefficients in a solar-like convective dynamo solution
with irregular long term behaviour. We have particularly focussed on two
special properties
seen in the dynamo solution, namely the erratic-looking parity switches
(competition between dipolar and quadrupolar dynamo modes), and
the abrupt disappearance and re-emergence of surface magnetic
activity. 

Our analysis revealed that the parity switches may be linked to variations seen
in the parity of the turbulent transport coefficients themselves, especially
related to the $\alpha_{rr}$ component near the bottom of the
convection zone.
We also found an indication, that during a 'quadrupole epoch', the region of negative
shear in the mid-latitudes, a prominent feature present in many convection simulations
and shown to be responsible for the equatorward migration
\citep{WKKB14}, becomes suppressed, and this may lead to more
chaotic-appearing migration patterns.

During the `missing cycle', we observed a disappearance of surface magnetic activity lasting roughly
 one magnetic cycle of five years in the northern hemisphere.
Simultaneously, we observed intensification of the magnetic
field at the base of the convection zone. Although this epoch does not really represent a grand minimum,
this behavior is very similar to that found by \cite{KKOBWK16} from their simulation exhibiting a
more extended minimum epoch. 
Our test-field analysis provides a plausible explanation of it in terms of the
reduction of $\alpha_{\phi\phi}$ prior to the reduced surface
activity, and enhanced turbulent downward pumping during the event confining some of the magnetic field at the bottom
of the convection zone. 
At the same time, however,
a quenching of the turbulent magnetic diffusivities is observed,
albeit distributed somewhat differently by depth compared to 
the $\aalpha$ components.
However, to determine the effect of these modifications to the turbulent transport
coefficients on the dynamo efficiency, dedicated mean-field modelling is required.

\acknowledgements
We gratefully acknowledge the input of Axel Brandenburg, Petri
J. K\"apyl\"a,
and Matthias Rheinhardt to this work.
Financial support from the Academy of Finland ReSoLVE Center of Excellence (grant
No. 272157; FAG, MJK) is acknowledged.
J.W.\ acknowledges funding from the People Programme (Marie Curie
Actions) of the European Union's Seventh Framework Programme
(FP7/2007-2013) under REA grant agreement No.\ 623609.
The simulations were performed using the supercomputers hosted by the
CSC – IT Center for Science Ltd. in Espoo, Finland, which is administered by
the Finnish Ministry of Education.

\bibliography{refs}
\bibliographystyle{an}
\end{document}